\documentclass[manuscript]{aastex}

\shorttitle{Faint Companions}
\shortauthors{Chakraborty et al.}

\begin{document}


\title{Nature of Faint Companions to G type Stars using Adaptive Optics.}


\author{A. Chakraborty\altaffilmark{}, J. Ge\altaffilmark{} \and 
J.H. Debes\altaffilmark{}}
\affil{Pennsylvania State University, University Park, PA 16802}
\email{abhijit@astro.psu.edu, jian@astro.psu.edu, debes@astro.psu.edu}




\begin{abstract}
High spatial resolution (0.30 arcsec) NIR photometric observations using the adaptive 
optics of the 100 inch Mt. Wilson Telescope and the Penn State IR Imager and 
Spectrograph (PIRIS) have revealed faint companions to HD190067 and HIP13855. HD190067b 
is found to be of mass of 0.075$M_{\odot}$ to 0.1$M_{\odot}$ and with an age of 0.5 Gyrs 
to a few Gyrs. HIP13855b is a low mass star (age: 100 Myrs to 500 Myrs) of mass range between 
0.1$M_{\odot}$ to 0.2$M_{\odot}$. Further NIR spectroscopic observations will be necessary 
to classify their spectral type precisely.   
\end{abstract}




\section{Introduction}
Substellar dwarfs or brown dwarfs are objects whose masses are too low to burn 
hydrogen at their cores. The mass could range between 0.075M${_\odot}$  to 
0.002M${_\odot}$ (late M, L and T type stars, see for eg: Kirkpatrick \& McCarthy 
1994, Burrows et. al. 1997, Kirkpatrick et. al. 1999a, Basri 2000). In the last few years 
ultracool dwarfs have discovered from the 2MASS, Sloan and DENIS surveys (Kirkpatrick 
et al. 1999a, 2000, Strauss et al. 1999, Fan et al. 2000, and Delfosse et al. 1997). 
These extensive works helped in establishing spectral signatures of ultracool dwarfs
and classifications of M, L and T spectral types (also see Reid et al. 2001 and Leggett 
et al. 2002).

However, determining the intrinsic properties of brown dwarfs is difficult because their 
luminosity and effective temperatures are function of age and mass (Burrows et al. 
1997, Baraffe et al. 1998). Therefore, finding such objects in binary systems containing 
a main sequence star would be of great interest because both these parameters can be 
determined reasonably accurately. Indeed, the first discovered undisputed brown dwarf is 
a companion to M1V star GL229 (Nakajima et al. 1995). Also, the coronographic Palomar 
survey (Oppenheimer et al. 2001) which goes 4 absolute magnitudes fainter than GL229b, 
discovered at least seven new faint companions of nearby stars.  

We have started a high angular resolution (= 0\farcs30) search for faint companions at 
NIR wavelengths around G or K type stars in the solar neighborhood, as well as in the 
nearby young star clusters like MBM12 and MBM20 (Hearty et al. 2000a, 2000b). The list
consisted of stars with ages from 100 Myrs to a few Gyrs using ground based telescope 
equipped with adaptive optics. In this paper we report the discovery of two faint companions 
to main sequence stars: HD190067 and HIP13855. This is a result of our first limited survey 
done using the 100 inch (2.5m) telescope of Mt.Wilson. 

\section{Observations and Data Analysis}
The H and K photometric observations were performed in October 2001 at the 100 inch 
telescope of Mt.Wilson using its natural star Adaptive Optics (AO) system (Shelton et 
al. 1995) and the Penn state IR Imager and Spectrograph (PIRIS, Ge et al. 2002a). The 
detector is a 256$\times$256 PICNIC array with pixel size of 40 microns. The gain and 
the read noise of the camera are 4.0 $e^{-}/ADU$ and 20 electrons respectively. The 
plate scale was 0\farcs082 per pixel providing a field of view of 21\farcs0. The 
filters used were standard Astronomical H \& K filters. A cold pupil mask located in 
the pupil plane inside the dewar reduces the thermal background particularly in the K 
band. The camera will soon be equipped for NIR spectroscopy with $R=200$ by commercially 
available fused silica grisms and high $R=5000$ using silicon grisms developed at Penn 
State (Ge et al. 2000, Ge et al. 2002b). 

We observed six young solar analogs from Gaidos et al. (2000), a couple of main sequence 
stars with faint companions (Turner et al. 2001) and six main sequence stars towards the 
MBM12 star forming cloud (Hearty et al. 2000a\&b). The observations were done under 
photometric condition with sub arcsec seeing. The average FWHM of the Point Spread Function 
(PSF) was 0\farcs30 after the AO correction. 

We report here the observations of two stars HD190067 (spectral type G8, Duquennoy \& 
Mayor 1991) and HIP13855 (spectral type G0, Hearty et al. 2000b) which are associated 
with faint companions (HD190067b and HIP13855b). We discuss the nature of these faint 
companions and determine their possibility of being substellar, transition or late M
dwrafs close to the hydrogen burning limit.

The total integration in $H$ and $K$ bands for HD190067 was of 20 s (5 co-added 4 s 
exposures) and 50 s (5 co-added 10 s exposures) for HIP13855. An appropriate number of 
sky frames (5) were observed immediately after the object observations for proper sky 
subtraction. Flats were made by combining several (15) sky frames. Two standard stars 
HD225023 and SAO056596, were observed from the UKIRT standard star list and Hunt et al. 
(1998) respectively for determining the zero point of the Camera in each filter before 
or after the observation of the program stars. Each frame was sky subtracted and flat 
fielded before co-adding. Photometry was performed on both individual frames as well as 
in the total co-added frames to estimate the total photometric error. Photometric 
analysis was carried out using the standard DAOPHOT routine of IRAF. While primarily we 
did aperture photometry after determining the width of the PSF using PHOT task, we also 
performed photometry of the faint companion HIP13855B after subtracting the PSF fitted 
bright primary HIP13855A using PSF and ALLSTAR tasks. This procedure was necessary 
because the separation between A and B components of HIP13855 is only 0\farcs5. A 
Gaussian PSF was constructed considering an isolated bright star. The photon noise 
correspond to an error of $\pm$0.02 mag in the individual images. However, we estimated the 
maximum error by computing the magnitudes from five different frames to be $\pm$0.1 mag and 
the mean magnitude from five frames is the same as the magnitude estimated from the 
co-added frames. This relatively high error could be due to the AO system and in the 
present work we will limit the uncertainty to $\pm$0.1 mag. Table 1 gives the value of 
magnitudes estimated from the co-added frames.  

We note that the source seen as a faint companion to HIP13855 with a separation of 
0\farcs5 is real and not an artifact like ghost. Other stars observed under similar 
conditions including the standard stars do not show any such point source which is 5 
magnitude fainter than the primary. 

We compared our $H$ and $K$ photometry of HIP13855A with that given in the 2MASS point 
source catalogue. The 2MASS H and Ks magnitudes of HIP13855 (A+B) is 7.29 and 7.25 
respectively. We find that if we consider a larger aperture (3 arcsec) comparable to 
the 2MASS PSF, then we do get similar values in H and K bands (7.4 and 7.3 respectively 
) within the our photometric errors stated earlier. 
    
\section{Results and Discussion}
The photometric results of HD190067, HIP13855 and their respective companions
are presented in Table 1. The absolute magnitudes of the faint companions are 
calculated assuming the distances given in the Hipparcos catalogue. A$_H$ values 
towards these sources from the literature are negligible compared to the overall 
photometric error ($\pm$0.1 mag) and hence are not considered for absolute magnitude 
estimation. The $A_H$ values were estimated from the $A_V$ values found in the 
literature for HIP13855 (Luhman 2001); for HD190067 mean extinction in the Galaxy was 
considered for a distance of 19 pc (Glass 1999), and using the relation $A_H/A_V$ = 
0.142 (Rieke \& Lebofsky 1985). Thus the estimated $A_H$ values for HIP13855 and 
HD190067 were 0.028 and 0.01 magnitudes. Figures 1 \& 2 show H and K images of HD190067 
and HIP13855 along with their faint companions respectively. 

\subsection{Ages of HD190067 and HIP13855}  
HD190067 is a disk star (Eggen 1987), and based on its velocity components in the disk it 
seems that its age could be anything between 0.5 Gyrs to a few Gyrs (Eggen 1987, Mihilas 
\& Binney 1981). 

HIP13855 is at a distance of 74 pc (Hipparcos catalogue) a foreground star towards the 
star forming cloud MBM12, whose distance is estimated to be somewhere between 90 pc to 
275 pc (Luhman 2001). MBM 12 cloud contain very young stars with ages 10 to 100 Myrs 
(Hearty et al. 2000b). Some members of the MBM12 cluster, however, are much closer; for 
example, HD17332 is at a distance of 33 pc (Hipparcos catalogue). HD17332 is a young 
G1V star, since lithium is detected in its spectrum (Hearty et al. 2000b). We 
qualitatively argue on the basis of its proximity that HIP13855 could be associated 
with the MBM12 cluster; however, there is no record of lithium detection in its 
spectrum. So we assume a lower limit for the age of HIP13855 could be 100 Myrs (Hearty et al. 
2000b).  
 
Burrows et al. (1997) developed models for young substellar dwarfs of various ages that 
show a similar trend of absolute magnitudes vs NIR infrared colors for young substellar 
dwarfs that are less than 1.0 Gyrs. Similar models of evolution of brown dwarfs and low 
mass stars with a time scale from less than 100 Myrs to 10 Gyrs have shown that the 
hydrogen burning limit is somewhere between 0.072 to 0.075$M_{\odot}$ and the 
corresponding spectral type varies with age (M6.5 at 100 Myrs to L4 at 10Gyrs (Baraffe et 
al. 1998 and 2001, Kirkpatrick 1999b, Basri 2000). Figure 3 shows such evolution model 
curves from Baraffe et al. (1998 and 2001) and overlaying on it are the results of the 
present work. We discuss below the nature of the newly discovered companion based on 
their ages and absolute $H$ magnitude. 

\subsection{HD190067}
HD190067 is at a distance of 19 pc (Hipparcos catalogue). The possible binary nature 
was first reported by Turner et al. (2001). However, they could not confirm it due to 
lack of information on common proper motion. We detect the binary component HD190067b 
at a separation of 2\farcs90 which is 0\farcs04 more than previous observed separation 
in 1996 by Turner et al. (2001). HD190067 should have moved by 0\farcs51 between these 
two set of observations due to its proper motion alone. Thus, it appears that HD190067 
and its faint companion has similar proper motion and establish that HD190067b is a 
physical binary. However, the observed PA in 1996 was 82 degrees while in 2001 (present 
work) we find it to be 75 degrees (see Table 1). This difference could be due to the 
orbital motion of the secondary. We will need more such observations to determine the 
orbital period and the mass of the secondary.  

We estimate the absolute H magnitude M$_{H}$ = 9.3$\pm$0.1 and a $H-K$ color of 1.1$\pm$0.2. 
Kirkpatrick \& McCarthy (1994) showed from $R,I,J,H,K$ photometry and optical 
spectroscopy that young stars (less than 0.5 Gyrs) of spectral type M6.5 and later cannot 
burn hydrogen in their core. These measurements suggest that HD190067b could be a transition 
candidate if it is young (less than 0.5 Gyrs). However, if it is older, then it is a low 
mass star close to the hydrogen burning limit. From Figure 3 the mass of HD190067b may be 
0.08$M_{\odot}$ to 0.1$M_{\odot}$(also see Figure 3). The high H-K color (1.1$\pm$0.2) is 
intriguing. A high H-K color is common for late L type substellar dwarfs Reid et al. (2001), 
but a late L type source is expected to be more than a magnitude fainter in $M_H$ (Kirkpatrick 
et al. 2000). NIR spectroscopic observations will be necessary to accurately determine the 
spectral type. 

\subsection{HIP13855}
We find a faint companion of HIP13855 with a separation of 0\farcs5 whose H and K 
magnitudes are 12.6 and 12.6 respectively. These yield $M_{H}$ to be 8.3$\pm$0.1 for the 
HIP13855b. Figure 3 shows that the companion could have mass in the range of 0.1$M_{\odot}$ 
to 0.2$M_{\odot}$. We find the $H-K$ color of HIP13855b to be 0.0$\pm$0.2 mag. Therefore, 
if the system is a physical binary (common proper motion yet to be confirmed) then HIP13855b 
is a young late M dwarf. Spectroscopic observations will be helpful to probe further into its 
nature.  

\acknowledgments
We thank the Mt. Wilson telescope staff and the management for their help during the 
observations. We thank Dr. D. McCarthy for loaning us part of optics for PIRIS, Dr. R. 
Brown for the PICNIC array and Dr. A. Kutyrev for the filters. We also thank Profs. 
Larry Ramsay, Eric Feigelson, and Donald Schneider of Pennsylvania State University for 
their comments on the present work which significantly helped to improve the quality. 
The work was supported by the following grants from NASA: NAG5-10617, NAG5-11427 and 
NASA GSRP fellowship (Award No. NGT5-119) and Penn State Eberly College of Science.

\clearpage
\begin{figure}
\plottwo{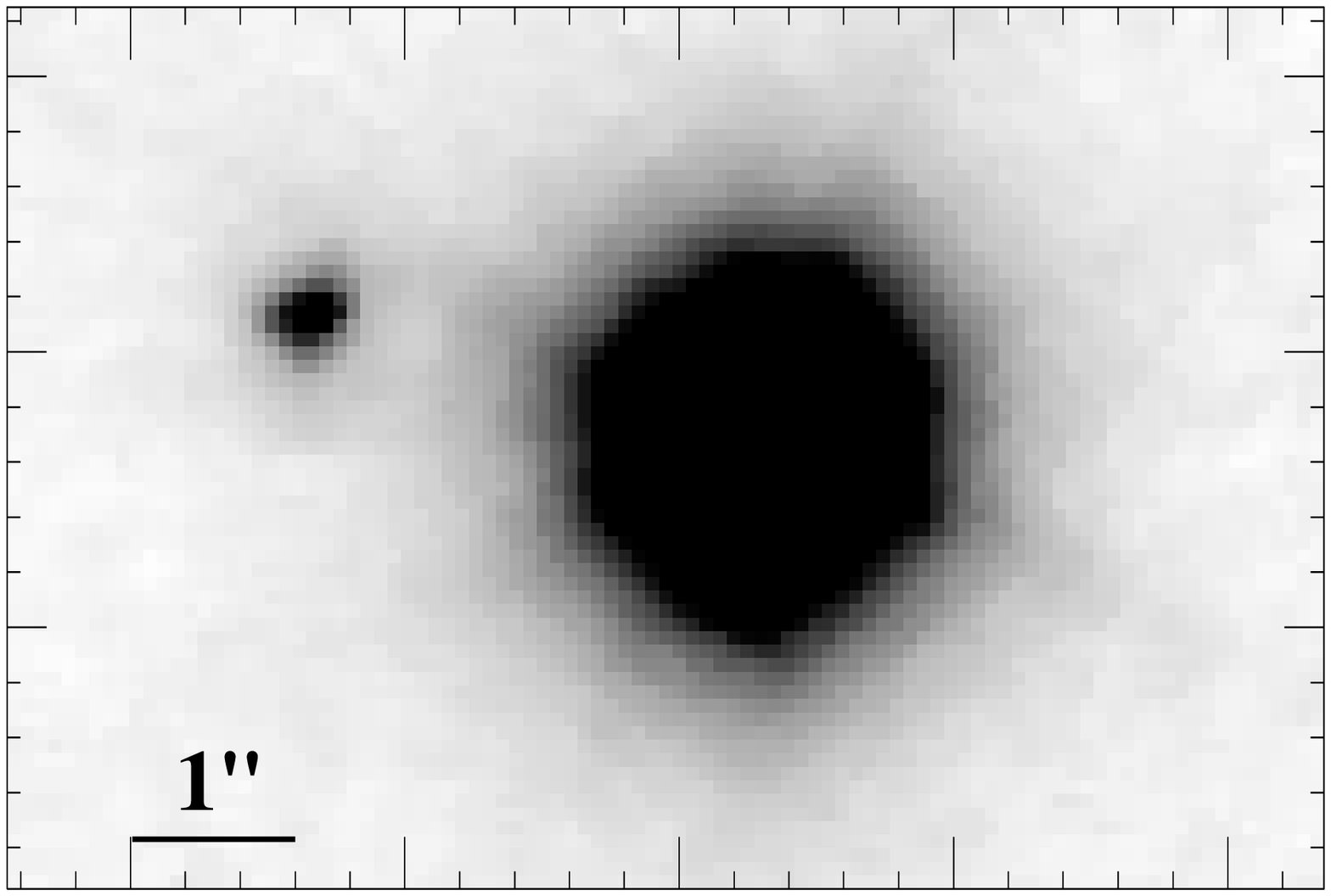}{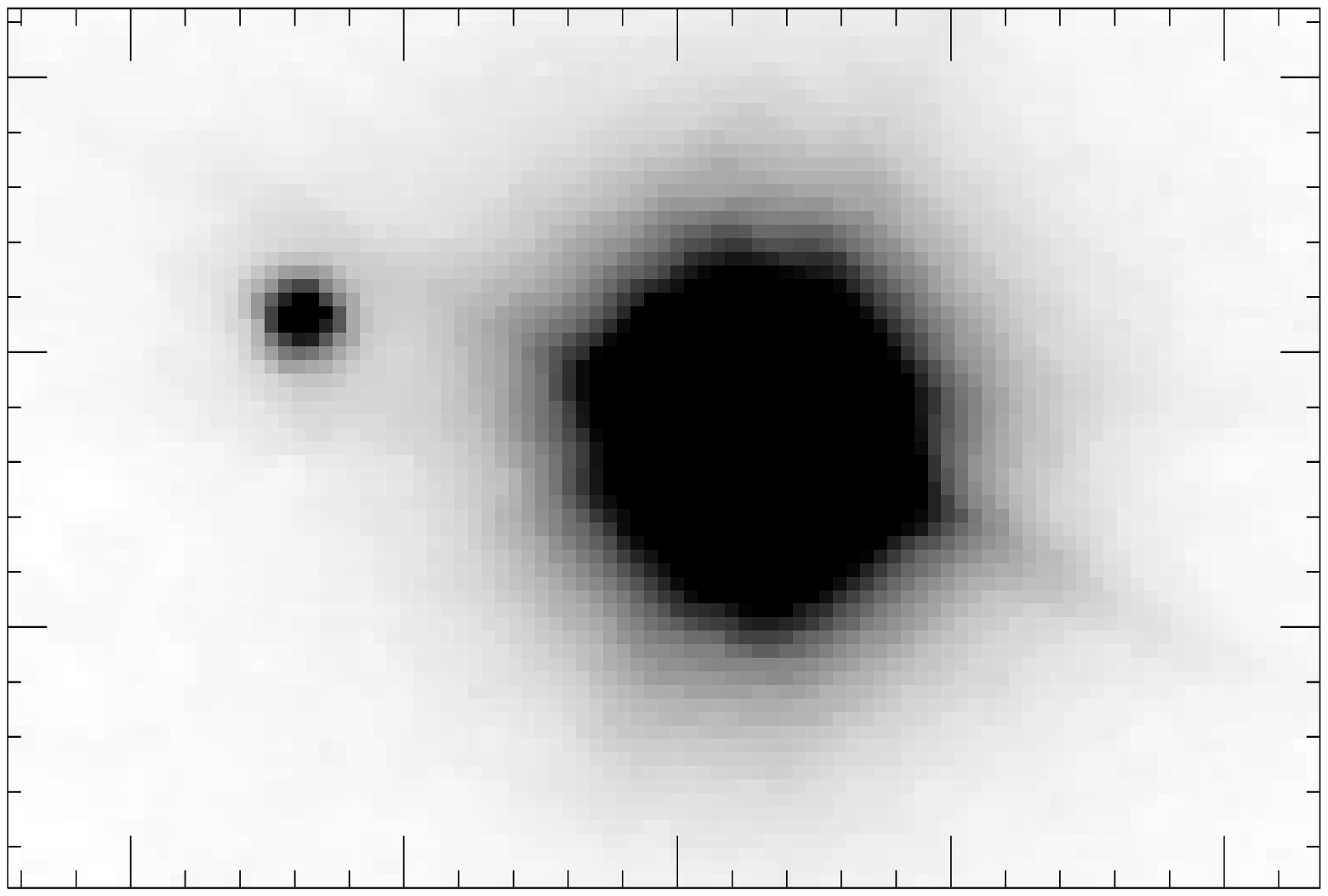}
\caption{H and K images of HD190067. 
North is up and East is to the Left.\label{fig1}}
\end{figure}

\clearpage
\begin{figure}
\plottwo{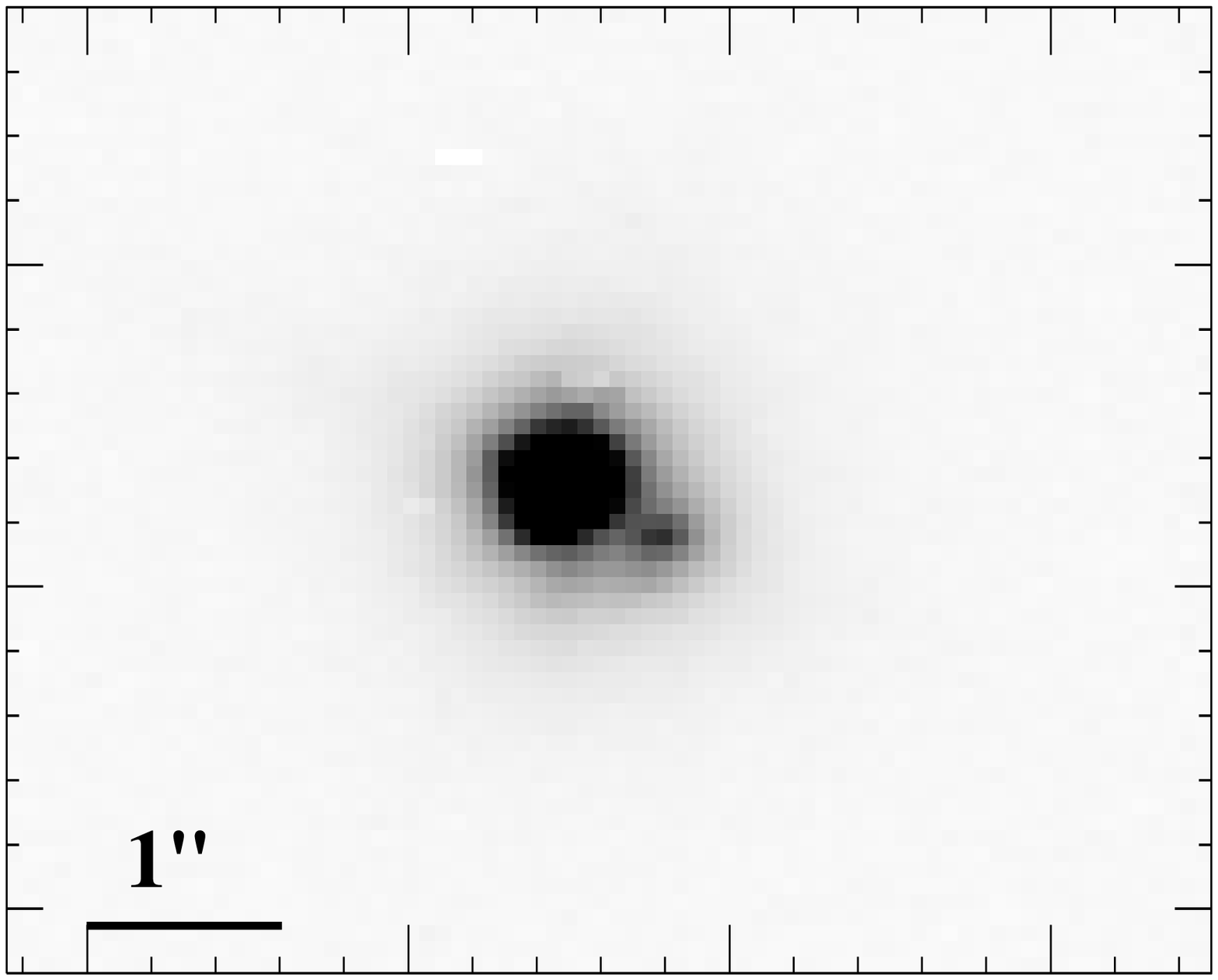}{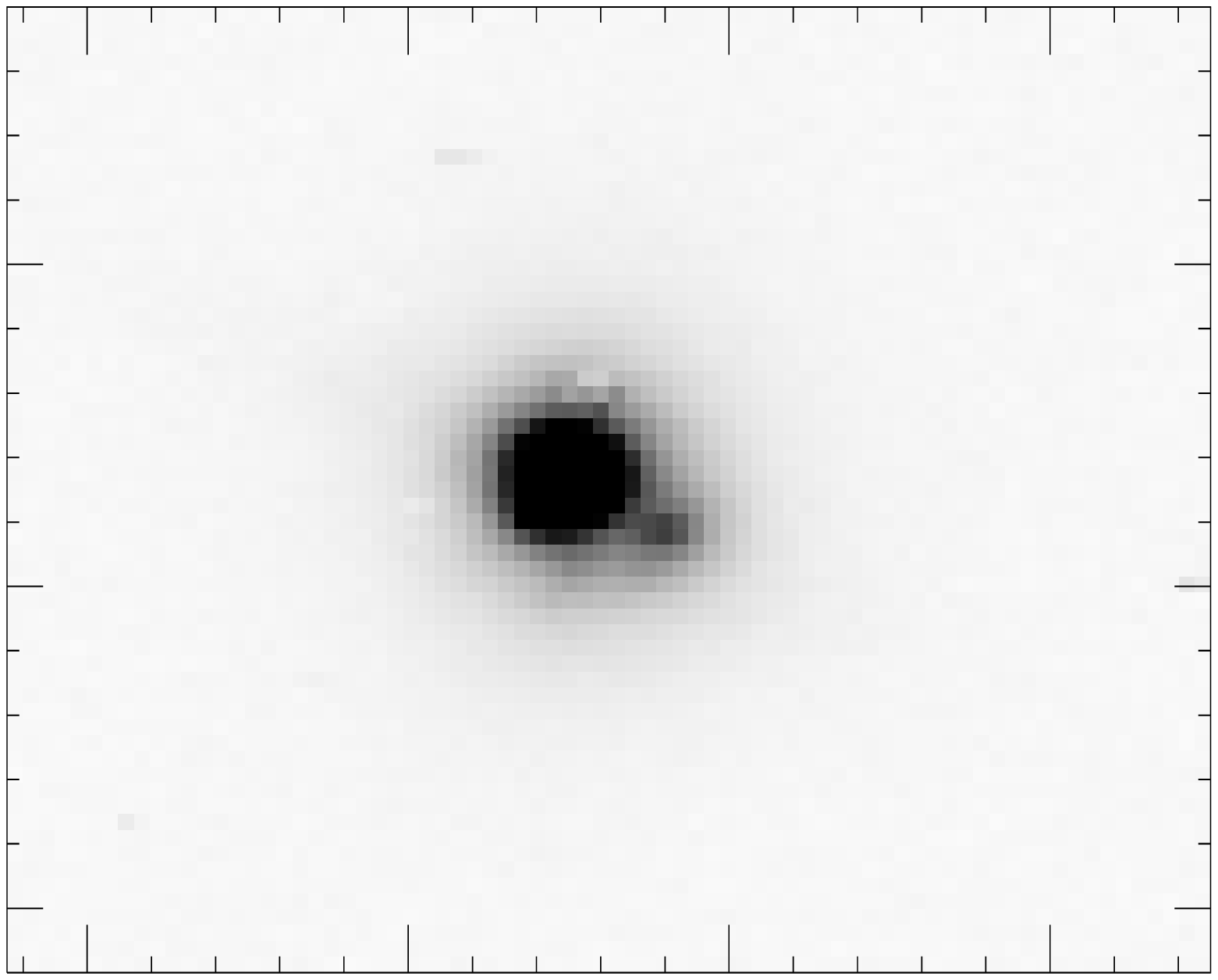}
\caption{H and K images of HIP13855. 
North is up and East is to the Left.\label{fig2}}
\end{figure}

\clearpage
\begin{figure}
\plotone{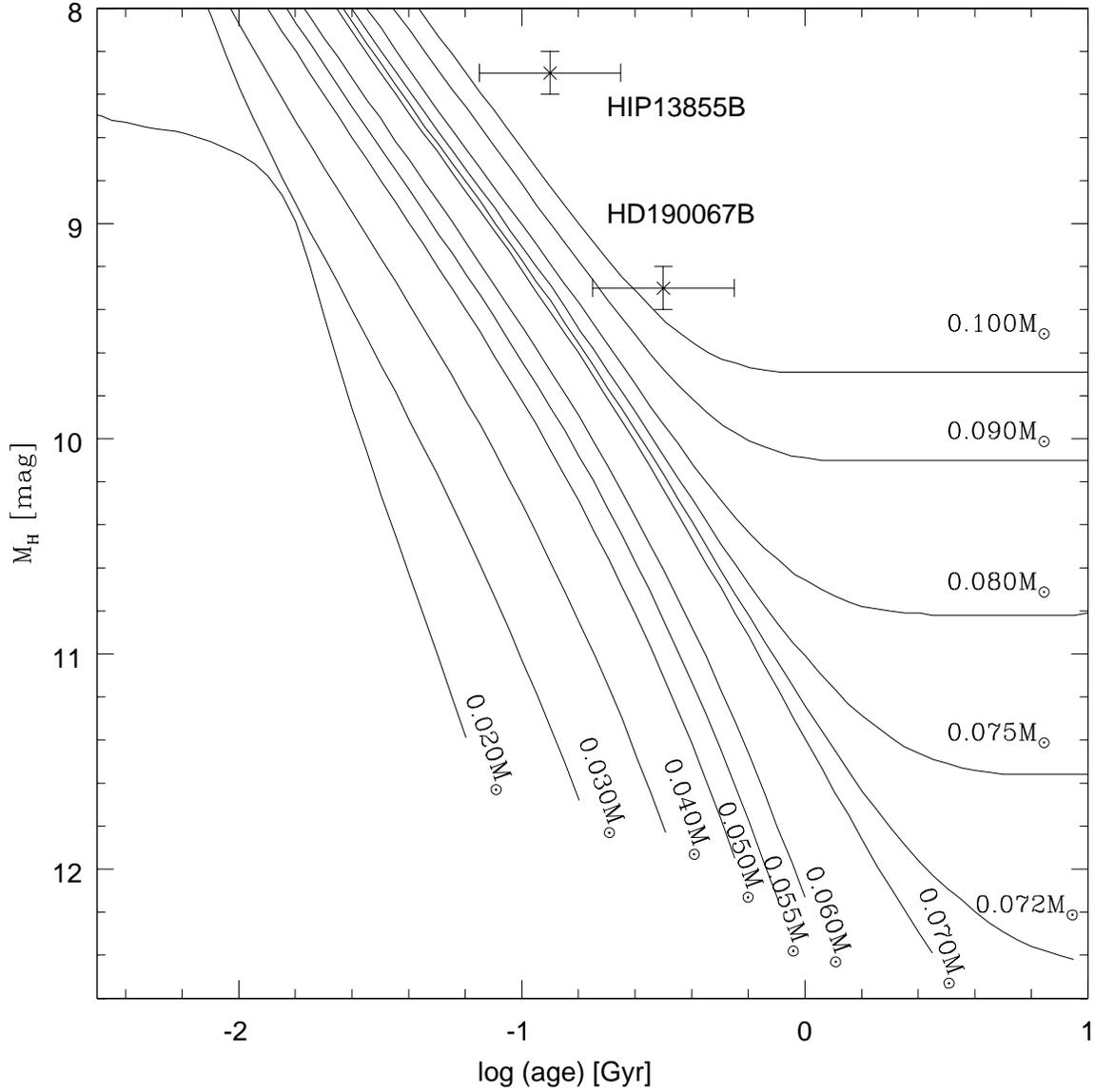}
\caption{Evolution of substellar dwarfs: absolute H magnitude vs Age (Gyrs). The solid 
lines are evolutionary model calculations of low mass stars and Brown Dwarfs from Baraffe 
et al. (1998 and 2001) with metalicity $[M/H]=0$. The transition from substellar to 
stellar is somewhere between  0.072 to 0.075$M_{\odot}$ (Kirkpatrick 1999b, Basri 2000). 
See text for details. \label{fig3}}
\end{figure}

\clearpage
\begin{table}
\begin{center}
\caption{Photometric Properties of the Observed Objects. The overall photometric error 
in each band is $\pm$0.1 mag as discussed in section 2. The error in the PA is about $\pm$0.4 
degrees.\label{tbl-1}}
\begin{tabular}{ccccc}
\tableline\tableline
Name & H(mag) & K(mag) & separation (arcsecs) & PA (degrees)\\
\tableline
HD190067A & 6.6& 6.4&&\\ 
HD190067B & 10.7 & 9.6& 2.9& 75\\
HIP13855A & 7.8 & 7.7&&\\ 
HIP13855B & 12.6 & 12.6& 0.5& 243\\
\tableline
\end{tabular}
\end{center}
\end{table}

\end{document}